\documentclass[
    ,final            
  ]
  {aipproc}

\layoutstyle{8x11double}

\newcommand{\xte}{{\it RXTE}}

\newcommand{\tfe}{1E~1048.1--5937}
\newcommand{\tfn}{1E~2259$+$586}

\newcommand{\oft}{4U~0142$+$61}

\newcommand{\axj}{AX~J1845--0258}
\newcommand{\ett}{XTE~J1810--197}

\begin{document}

\title{Activity From Magnetar Candidate 4U~0142+61: Bursts and Emission Lines}



\author{Fotis~P.~Gavriil}{
  address={NASA Goddard Space Flight Center, Astrophysics Science
                 Division, Code 662, Greenbelt, MD, 20771, USA},
altaddress={CRESST; University of Maryland Baltimore County, Baltimore,
                 MD, 21250, USA}
}

\author{Rim~Dib}{
  address={Department of Physics, McGill University,
                 Montreal, QC, H3A~2T8, Canada}
}

\author{Victoria~M.~Kaspi}{
  address={Department of Physics, McGill University,
                 Montreal, QC, H3A~2T8, Canada}
}

\begin{abstract}
After 6 years of quiescence, Anomalous X-ray Pulsar (AXP) 4U~0142+61
entered an active phase in 2006 March that lasted several
months. During the active phase, several bursts were detected, and
many aspects of the X-ray emission changed.  We report on the
discovery of six X-ray bursts, the first ever seen from this AXP in
$\sim$10 years of \textit{Rossi X-ray Timing Explorer} monitoring. All
the bursts occurred in the interval between 2006 April~6 and 2007
February~7.  The burst durations ranged from 8$-$3$\times$10$^{3}$~s
as characterized by $T_{90}$. These are very long durations even when
compared to the broad $T_{90}$ distributions of other bursts from AXPs
and Soft Gamma Repeaters (SGRs).  The first five burst spectra are
well modeled by simple blackbodies, with temperature
$kT\sim2-6$~keV. However, the sixth and most energetic burst had a
complicated spectrum consisting of at least three emission lines with
possible additional emission and absorption lines.  The most
significant feature was at $\sim14$~keV. Similar 14-keV spectral
features were seen in bursts from AXPs \tfe\ and \ett. If this feature
is interpreted as a proton cyclotron line, then it supports the
existence of a magnetar-strength field for these AXPs.  Several of the
bursts were accompanied by a short-term pulsed flux enhancement.  We
discuss these events in the context of the magnetar model.
\end{abstract}

\maketitle

\section{Introduction}

Anomalous X-ray Pulsars (AXPs) are isolated neutron stars that show
pulsations in the narrow range of 2--12~s. Their observed 2-10~keV
X-ray luminosities ($\sim 10^{33}-10^{35}$~erg~s$^{-1}$) cannot be
accounted for by their available spin-down energy.  It is widely
accepted that AXPs are magnetars -- young
isolated neutron stars powered by their high magnetic fields
\citep{td95, td96a}. The inferred surface dipolar magnetic fields of
AXPs are all above 5.9$\times$10$^{13}$~G.  The magnetar model was
first proposed to explain Soft Gamma Repeaters (SGRs). SGRs show
persistent properties similar to AXPs, but they were first discovered
by their enormous bursts of soft gamma rays ($> 10^{44}$~erg) and
their much more frequent, shorter, and thus less energetic bursts of
hard X-rays. To date, SGR-like X-ray bursts have been observed from
four AXPs, thus solidifying the connection between the two source
classes \citep{gkw02,kgw+03,wkg+05,kbc+06}. For a review of magnetar
candidates see \citet{wt06}.

Thus far, only the magnetar model can explain the bursts observed from
SGRs and AXPs \citep{td95}. The internal magnetic field exerts
stresses on the crust which can lead to large scale rearrangements of
the external field, which we observe as giant flares. If the stress is
more localized, then it can fracture the crust and displace the
footpoints of the external magnetic field which results in short X-ray
bursts.  The highly twisted internal magnetic field also slowly twists
up the external field and it is believed that the magnetospheres of
magnetars are globally twisted \citep{tlk02}.  Reconnection in this
globally twisted magnetosphere has also been proposed as an additional
mechanism for the short bursts \citep{lyu02}.

In addition to bursts, AXPs and SGRs exhibit pulsed and persistent
flux variations on several timescales.  An hours-long increase in the
pulsed flux has been seen to follow a burst in AXP
\tfe\ \citep{gkw06}.  On longer timescales, AXPs can exhibit abrupt
increases in flux which decay on $\sim$week-month timescales. These
occur in conjunction with bursts and are thought to be due to thermal
radiation from the stellar surface after the deposition of heat from
bursts. Such flux enhancements have been observed in SGRs (see
\citet{wkg+01} for example). The flux enhancement of AXP \tfn\ during
its 2002 outburst was also interpreted as burst afterglow
\citep{wkt+04}, however, a magnetospheric interpretation has also been
proposed \citep{zkw+07}.  AXP \tfe\ exhibited three unusual flux
flares. In the first two, the pulsed flux rose on week-long timescales
and subsequently decayed back on time scales of months
\citep{gk04,tgd+07}. Although small bursts sometimes occur during
these events \citep{gkw06}, burst afterglow cannot explain the
flaring, thus these variations have been attributed to twists
implanted in the external magnetosphere from stresses on the crust
imposed by the internal magnetic field.  AXPs \ett\ and the AXP
candidate \axj\ have also exhibited flux variations, however it is not
clear whether these were of the abrupt rise type as in \tfn\ or the
slow-rise type as in \tfe.  Finally, AXP \oft\ has exhibited the
longest timescale flux variations, in which the pulsed flux increased
by 19$\pm$9\% over a period of 2.6 years \citep{dkg07}.

\section{Analysis and Results} 

All data presented here are from the Proportional Counter Array (PCA)
aboard the \textit{Rossi X-ray Timing Explorer} (\xte). The PCA is
made up of five identical and independent proportional counter units
(PCUs). Each PCU is a Xenon/methane proportional counter with a
propane veto layer. The data were collected in either
\texttt{GoodXenonwithPropane} or \texttt{GoodXenon} mode which record
photon arrival times with $\sim$1-$\mu$s resolution and bins them with
256 spectral channels in the $\sim$2-60~keV band.

\subsection{Burst Analysis}
\label{sec:bursts}

We have been monitoring \oft\ with the PCA for nearly a
decade. Currently, it is observed bi-monthly with a typical observation
length of 5~ks.  For each monitoring observation of 4U~0142+61, using
software that can handle the raw telemetry data, we generated 31.25~ms
lightcurves using all Xenon layers and only events in the 2--20~keV
band. These lightcurves were searched for bursts using our automated
burst search algorithm introduced in \citet{gkw02} and discussed
further in \citet{gkw04}. In an observation on 2006 April 6, we
detected a significant burst, and four more bursts were detected in a
single observation on 2006 June 25. The sixth and most energetic burst
was detected on 2007 February 7. There were 3, 3, and 2 PCUs on at the
times of the bursts for the April, June and February observations,
respectively. The bursts were significant in each active PCU.

\begin{figure}
\includegraphics[width=.70\columnwidth]{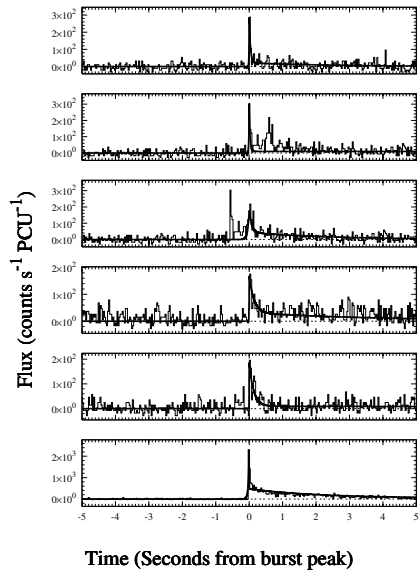} 
\caption { The histograms are the 2--60~keV burst lightcurves binned
  with 1/32~s resolution as observed by \xte. The thick curves are the
  best fit exponential rise and exponential decay model.
\label{fig:lc}
}
\end{figure}

To further analyze these bursts we created event lists in
FITS\footnote{\url{http://fits.gsfc.nasa.gov}} format using the
standard
\texttt{FTOOLS}\footnote{\url{http://heasarc.gsfc.nasa.gov/docs/software/ftools/}}.
For consistency with previous analysis of SGR/AXP bursts we extracted
events in the 2-60~keV band. These events were barycentered using the
position found by \citet{pkw+03} for the source.  The burst
lightcurves are displayed in Fig.~\ref{fig:lc}.

Before measuring any burst parameters we determined the instrumental
background using the \texttt{FTOOL} \texttt{pcabackest}. We extracted
a background model lightcurve using the appropriate energy band and
number of PCUs. \texttt{pcabackest} only determines the background on
16~s time intervals, so we interpolated these values by fitting a
polynomial of order 6 to the entire observation, which yielded a good
fit for each observation.

The burst peak time, rise time and peak flux were determined using the
methods described in \citet{gkw04}.  Usually, to measure the fluence
for SGR and AXP bursts, we subtract the instrumental background for
the lightcurve, integrate the light curve and fit it to a step
function with a linear term whose slope is the ``local'' background
rate. The fluence in this case is the height of the step
function. Although this technique worked well for the first burst,
which was a short isolated event, it was not appropriate for bursts 2,
3, and 4 because they had overlapping tails, and bursts 5 and 6 had
tails that extended beyond the end of the observation. Thus, we opted
to fit the bursts to exponential rises with decaying tails. Our model
fits are overplotted on the bursts in Fig.~\ref{fig:lc}. As is done
for $\gamma$-ray bursts and SGR and AXP bursts, we characterized the
burst duration by $T_{90}$, the time from when 5\% to 95\% of the
total burst counts have been collected. To determine the $T_{90}$
duration we integrated our burst model and numerically determined the
5\% and 95\% time crossings.  All burst temporal parameters are
presented in Table~\ref{tab:bursts}.

\begin{table}
\caption{Burst Temporal and Spectral Properties \label{tab:bursts}}
\begin{tabular}{lcccccccc}
\hline\hline
               & \multicolumn{1}{c}{April 2006} && \multicolumn{4}{c}{June 2006} && \multicolumn{1}{c}{February 2007} \\
\cline{2-2} \cline{4-7} \cline{9-9}
\tablehead{1}{c}{c}{Parameter\tablenote{All quoted errors represent 1-$\sigma$ uncertainties.}} & \tablehead{1}{c}{c}{Burst 1}              && \tablehead{1}{c}{c}{Burst 2} & \tablehead{1}{c}{c}{Burst 3} & \tablehead{1}{c}{c}{Burst 4} &  \tablehead{1}{c}{c}{Burst 5} && \tablehead{1}{c}{c}{Burst 6}  \\
Burst day (MJD UTC)                                          & 53831              && 53911   & 53911 & 53911  & 53911  &&  54138  \\
Burst start time (UT)     &  07:09.55.544(7)   &&   01:15:54.555(11)  &  01:15:55.119(43) &   01:16:9.216(4)     &  01:20:0.131(3)      && 10:04:43.264(27)   \\
Burst rise time, $t_r$ (ms)                                  & 7$^{+4}_{-3}$       && 11$^{+4}_{-3}$ & 43$^{+22}_{-22}$ & 4$^{+3}_{-3}$  & 3$^{+3}_{-1}$     &&  27$^{+4}_{-3}$\\
Burst duration, $T_{90}$ (s)                                 &  8.0$^{+2.5}_{-2.2}$ && 562.2$^{+0.15}_{-0.15}$ & 6.44$^{0.15}_{-0.15}$ & 11.62$^{+0.15}_{-0.15}$ & 92.54$^{0.70}_{-0.85}$ && 3472.9$^{3.3}_{-3.4}$\\         
$T_{90}$ Fluence (counts~PCU$^{-1})$                          & 65$\pm$6 && 1847$\pm$4 &125$\pm$9 & 147$\pm$8 & 342$\pm$9  && 14377$\pm$41\\
$T_{90}$ Fluence ($\times$10$^{-10}$erg~cm$^{-2}$)   & 16$\pm$3 && 229$\pm$11  & 26$\pm$3 & 38$\pm$3 & 43$\pm$4 && 1455$\pm$53\\
Peak Flux (counts~s$^{-1}$~PCU$^{-1}$)               & 299$\pm$59 && 367$\pm$65 & 250$\pm$54 & 228$\pm$52 & 245$\pm$54&& 551$\pm$97 \\
Peak Flux ($\times$10$^{-10}$ erg~s$^{-1}$~cm$^{-2}$)  & 75$\pm$19  && 46$\pm$8   & 53$\pm$12  & 58$\pm$14 & 31$\pm$7 && 253$\pm$22\\
Blackbody Temperature, $kT$\tablenote{For burst 6, the temperature is that of the blackbody component after accounting for the emission lines.} (keV)           & 6.2$^{+4.4}_{-1.9}$ && 2.67$^{+0.10}_{-0.09}$  & 4.50$^{+0.38}_{-0.29}$ & 5.16$^{+0.44}_{-0.34}$ & 2.52$^{+0.21}_{-0.13}$ && 2.12$^{+0.08}_{-0.04}$ \\
\hline
\end{tabular}
\end{table}

\begin{figure}
\includegraphics[scale=.62]{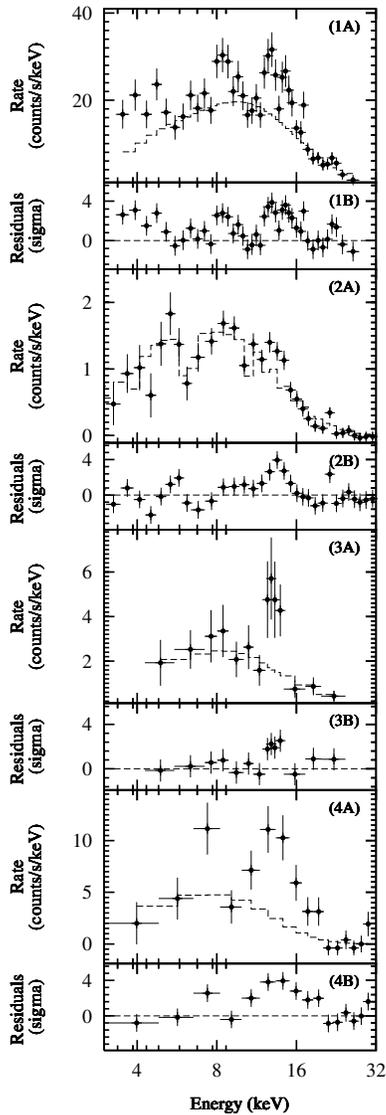}
\caption { Burst spectra of all AXP bursts with significant emission
  lines as observed by \xte. (1A) Burst spectrum of \oft\ burst 6. The
  dotted line indicates the continuum (blackbody) component of the
  best fit model. 1B: Residuals after subtracting the continuum
  component of the best fit model. 2A and 2B: Same but for burst 4 of
  \ett\ \citep[see][]{wkg+05}. 3A and 3B: Same, but for burst 3 of
  \tfe\ \citep[see][]{gkw06}. 4A and 4B: Same, but for burst 1 of
  \tfe\ \citep[see][]{gkw02}.
\label{fig:manysources}
}
\end{figure}

\begin{figure}
\includegraphics[width=0.9\columnwidth]{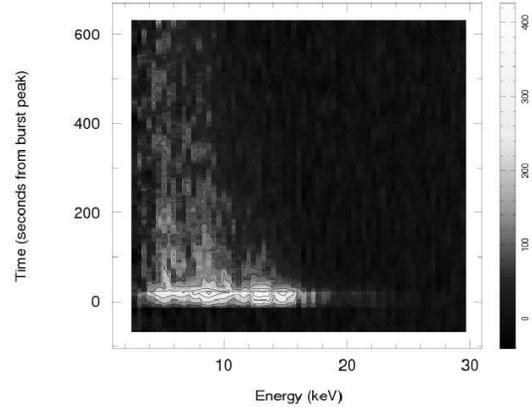}
\caption { Dynamic spectrum of burst 6. The wedge indicates the number
  of counts as a function of time and energy. Notice how at later
  times the contribution of the features at $\sim$4~keV and
  $\sim$8~keV exceed that of the 14~keV feature.
\label{fig:dynamic}
}
\end{figure}

Burst spectra were extracted using all the counts within their
$T_{90}$ interval. Background intervals were extracted from long,
hand-selected intervals prior to the bursts. Response matrices were
created using the \texttt{FTOOL} \texttt{pcarsp}. The burst spectra
were grouped such that there were at least 15 counts per bin after
background subtraction. Burst spectra, background spectra, and
response matrices were then read into the spectral fitting package
\texttt{XSPEC}\footnote{\url{http://xspec.gsfc.nasa.gov}} v12.3.1. The
spectra were fit to photoelectrically absorbed blackbodies using the
column density found by \citet{dv06b}. Only bins in the 2--30~keV band
were included in the fits. The blackbody model provided an adequate
fit for bursts 1 through 5. burst 6, however, was not well modeled by
any simple continuum model because of the presence of emission lines
(see Fig.\ref{fig:manysources} panels 1A and 1B). These features
showed clear temporal variability but they were most prominent near
the onset of the burst (see Fig.~\ref{fig:dynamic}).

\subsection{Pulsed Flux Analysis}
\label{sec:flux}

For each of the three observations containing bursts, we made two
barycentered time series in count rate per PCU, one for the 2$-$4~keV
band and the other for 4$-$20~keV. We only included the photons
detected by the PCUs that were on for the entire duration of the
observation. The time resolution was 1/32~s. We removed the 4~s
centered on each burst from each time series. Then, we broke each time
series into segments of length $\sim$500~s. For each segment, we
calculated the pulsed flux using two different methods.

\begin{figure}
\includegraphics[width=0.80\columnwidth]{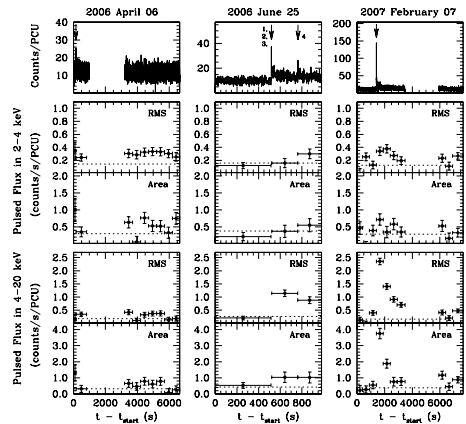}
\caption
{ RMS and area pulsed flux within the observations containing
  bursts. Each column corresponds to one observation. In each column
  we have, descending vertically, the 1-s resolution
  lightcurve with the bursts indicated, the 2--4~keV RMS pulsed flux,
  the 2--4~keV area pulsed flux, the 4--20~keV RMS pulsed flux, and the
  4--20~keV area pulsed flux. The dotted line in each of the pulsed
  flux plots shows the average of the pulsed fluxes obtained after
  segmenting and analyzing the time series of the observation
  immediately prior to the one shown.
\label{fig:fluxshort}
}
\end{figure}

First, we calculated the RMS pulsed flux using the Fourier
decomposition method described by \citet{wkt+04}, only incorporating
the contribution of the first 5 harmonics for consistency with
\cite{dkg07} and \cite{gdk+07}.  While least sensitive to noise, the
RMS method returns a pulsed flux number that is affected by pulse
profile changes (Archibald et al. in prep.). So to confirm our pulsed
flux results, we also used an area-based estimator to calculate the
pulsed flux, $PF = {a_0 - \frac{p_{\textrm{\small{min}}}}{N}}$, where
$a_0 = \frac{1}{N}{\sum_{i=1}^{N}} {p_i}$, $i$ refers to the phase
bin, $N$ is the total number of phase bins, $p_i$ is the count rate in
the $i^{\textrm{\small{th}}}$ phase bin of the pulse profile, and
$p_{\textrm{\small{min}}}$ is the average count rate in the off-pulse
phase bins of the profile, determined by cross-correlating with a high
signal-to-noise template, and calculated in the Fourier domain after
truncating the Fourier series to 5 harmonics.  The results are shown
in Figure~\ref{fig:fluxshort}.  Note the significant increase in the
4$-$20~keV pulsed flux in the 2007 June observation following the
cluster of bursts. This increase is not present in 2$-$4~keV. Also
note the significant rise and subsequent decay of the pulsed flux
following the large 2007 February burst.

\section{Discussion}
\label{sec:discussion}

We have discovered six bursts from AXP \oft. These bursts all occurred
between 2006 April and 2007 February, and were the only ones ever
observed from this source in $\sim$10 years of monitoring.  After the
first burst \oft\ exhibited a timing anomaly and pulse profile
variations (Gavriil, Dib, \& Kaspi in preparation). Together with the
short-term pulsed flux increase, the simultaneity of all these
phenomena clearly identifies \oft\ as the origin of the bursts.

\citet{wkg+05} first argued that there appear to be two classes of
magnetar bursts. Type A bursts are short, symmetric, and occur
uniformly in pulse phase. Type B bursts have long tails, thermal
spectra, and occur preferentially at pulse maximum.  \citet{wkg+05}
noted that type A bursts occur predominately in SGRs and type B bursts
occur predominately in AXPs, and this was affirmed by
\citet{gkw06}. Both argue that Type A and Type B bursts are produced
by different mechanisms. In the magnetar model bursts can either be
due to the rearrangement of magnetic field lines anchored to the
surface after a crustal fracture \citep{td95}, or due to reconnection
in the upper magnetosphere \citep{lyu02}. \citet{wkg+05} and
\citet{gkw06} argue that Type B bursts are due to the former and Type
A bursts are due to the latter.  The bursts reported here all had very
long tails, $T_{90}>8$~s, suggesting they are of Type B. However, the
bursts did not occur preferentially at pulse maxima.

The line-rich spectrum of burst 6 is intriguing. Three significant
features are seen at $\sim$4, $\sim$8 and $\sim$14~keV.  The most
significant emission feature at $\sim$14 keV is particularly
interesting. Emission features at similar energies were observed from
two out of the three bursts from \tfe\ \citep{gkw02,gkw06} and in one
out of the four bursts from \ett \citep{gkw02,wkg+05}. We have
reanalyzed these burst spectra in a consistent manner as for \oft. In
Fig.~\ref{fig:manysources} we plot the spectra of all AXP bursts with
emission lines in their spectra. Notice that all the spectra have
features that occur between 13 and 14~keV and are very broad. There is
clear evidence for features at $\sim$4 and $\sim$8~keV in the
\oft\ burst; however, note that there is subtle evidence for these
features in some of the other burst spectra as well.

If the $\sim$14~keV feature is interpreted as a proton cyclotron
feature then we can infer the surface magnetic field strength of the
star. For a line of energy $E$ the magnetic field strength is given by
\begin{equation}
B = \left(\frac{mc}{\hbar e}\right) E.
\label{eq:cyclotron}
\end{equation}
Setting $m$ equal to the proton mass we obtain $B = 2.2 \times 10^{15}
\left( {E}/{14~\mathrm{keV}} \right)~\mathrm{G}$. This field estimate
is much greater than that derived from the spin down of the source,
however the burst spectroscopic method measures the field at the
surface which can be multipolar, while the spin down measurement is
sensitive to the dipolar component.

The feature also could be an electron cyclotron feature at the
surface. Replacing $m$ with the electron mass in
Eq.~\ref{eq:cyclotron} we obtain $B =1.2\times10^{12} \left(
{E}/{14~\mathrm{keV}} \right)~\mathrm{G}$.  This field is two orders
of magnitude less than the spin down field.  However, if the feature
occurred higher up in the magnetosphere then the field would be
greatly reduced.  Thus, an electron cyclotron feature from a burst
which occurred in the upper magnetosphere cannot be precluded.

Although these features can, in principle, be proton/electron
cyclotron features there are many problems with interpreting them as
such.  First, it is not clear why three different sources, with
different magnetic field strengths would  exhibit features with
similar energies. Moreover, it is unclear why these features are
seldom seen and have not been seen in other high signal-to-noise
bursts.  Detailed modeling of burst spectra is definitely
warranted. The fact that these features occur at similar energies,
despite having different magnetic field strengths, may suggest that
they are actually atomic lines, and could possibly provide new
insights into the composition of the crust and atmosphere of these
enigmatic objects.

\bibliographystyle{aipproc}   

\begin{thebibliography}{20}
\expandafter\ifx\csname natexlab\endcsname\relax\def\natexlab#1{#1}\fi
\providecommand{\enquote}[1]{``#1''}
\expandafter\ifx\csname url\endcsname\relax
  \def\url#1{\texttt{#1}}\fi
\expandafter\ifx\csname urlprefix\endcsname\relax\def\urlprefix{URL }\fi
\providecommand{\eprint}[2][]{\url{#2}}

\bibitem[{Thompson} and {Duncan}(1995)]{td95}
C.~{Thompson}, and R.~C. {Duncan}, \emph{MNRAS} \textbf{275}, 255--300 (1995).

\bibitem[Thompson and Duncan(1996)]{td96a}
C.~Thompson, and R.~C. Duncan, \emph{ApJ} \textbf{473}, 322--342 (1996).

\bibitem[Gavriil et~al.(2002)]{gkw02}
F.~P. Gavriil, V.~M. Kaspi, and P.~M. Woods, \emph{Nature} \textbf{419},
  142--144 (2002).

\bibitem[Kaspi et~al.(2003)]{kgw+03}
V.~M. Kaspi, F.~P. Gavriil, P.~M. Woods, J.~B. Jensen, M.~S.~E. Roberts, and
  D.~Chakrabarty, \emph{ApJ} \textbf{588}, L93 (2003).

\bibitem[{Woods} et~al.(2005)]{wkg+05}
P.~M. {Woods}, C.~{Kouveliotou}, F.~P. {Gavriil}, {Kaspi}, V.~M., M.~S.~E.
  {Roberts}, A.~{Ibrahim}, C.~B. {Markwardt}, J.~H. {Swank}, and M.~H.
  {Finger}, \emph{ApJ} \textbf{629}, 985--997 (2005).

\bibitem[{Krimm} et~al.(2006)]{kbc+06}
H.~{Krimm}, S.~{Barthelmy}, S.~{Campana}, J.~{Cummings}, G.~{Israel},
  D.~{Palmer}, and A.~{Parsons}, \emph{The Astronomer's Telegram} \textbf{894},
  1--+ (2006).

\bibitem[Woods and Thompson(2006)]{wt06}
P.~M. Woods, and C.~Thompson, \enquote{{Soft Gamma Repeaters and Anomalous
  X-ray Pulsars: Magnetar Candidates},} in \emph{Compact Stellar X-ray
  Sources}, edited by W.~H.~G. Lewin, and M.~van~der Klis, Cambridge University
  Press, UK, 2006.

\bibitem[Thompson et~al.(2002)]{tlk02}
C.~Thompson, M.~Lyutikov, and S.~R. Kulkarni, \emph{ApJ} \textbf{574}, 332--355
  (2002).

\bibitem[Lyutikov(2002)]{lyu02}
M.~Lyutikov, \emph{ApJ} \textbf{580}, L65 (2002).

\bibitem[{Gavriil} et~al.(2006)]{gkw06}
F.~P. {Gavriil}, V.~M. {Kaspi}, and P.~M. {Woods}, \emph{ApJ} \textbf{641},
  418--426 (2006).

\bibitem[{Woods} et~al.(2001)]{wkg+01}
P.~M.~Woods, C.~Kouveliotou, E.~{G{\" o}{\u g}{\" u}{\c s}}, M.~H.~Finger, J.~Swank, D.~A.~Smith, K.~Hurley, C.~{Thompson}, \emph{ApJ} \textbf{552}, 748--755 (2001).

\bibitem[{Woods} et~al.(2004)]{wkt+04}
P.~M. {Woods}, V.~M. {Kaspi}, C.~{Thompson}, F.~P. {Gavriil}, H.~L. {Marshall},
  D.~{Chakrabarty}, K.~{Flanagan}, J.~{Heyl}, and L.~{Hernquist}, \emph{ApJ}
  \textbf{605}, 378--399 (2004).

\bibitem[{Zhu} et~al. (2007)]{zkw+07}
W.~{Zhu}, V.~M.~{Kaspi}, P.~M.~{Woods}, F.~P.~{Gavriil}, R.~{Dib}, \emph{ApJ} (2007), arXiv: 0710.1896 (astro-ph)

\bibitem[{Gavriil} and {Kaspi}(2004)]{gk04}
F.~P. {Gavriil}, and V.~M. {Kaspi}, \emph{ApJ} \textbf{609}, L67--L70 (2004).

\bibitem[{Tam} et~al.(2007)]{tgd+07}
C.~R. {Tam}, F.~P. {Gavriil}, R.~{Dib}, V.~M. {Kaspi}, P.~M. {Woods}, and
  C.~{Bassa}, \emph{ApJ}  (2007), arXiv:0707.2093(astro-ph).

\bibitem[{Dib} et~al.(2007)]{dkg07}
R.~{Dib}, V.~M. {Kaspi}, and F.~P. {Gavriil}, \emph{ApJ} \textbf{666},
  1152--1164 (2007).

\bibitem[Gavriil et~al.(2004)]{gkw04}
F.~P. Gavriil, V.~M. Kaspi, and P.~M. Woods, \emph{ApJ} \textbf{607}, 959--969
  (2004).

\bibitem[{Patel} et~al.(2003)]{pkw+03}
S.~K. {Patel}, C.~{Kouveliotou}, P.~M. {Woods}, A.~F. {Tennant}, M.~C.
  {Weisskopf}, M.~H. {Finger}, C.~A. {Wilson}, E.~{G{\"o}{\u g}{\"u}{\c s}},
  M.~{van der Klis}, and T.~{Belloni}, \emph{ApJ} \textbf{587}, 367--372
  (2003).

\bibitem[Durant and van Kerkwijk(2006)]{dv06b}
M.~Durant, and M.~H. van Kerkwijk, \emph{ApJ} \textbf{650}, 1082--1090 (2006).

\bibitem[{Gonzalez} et~al.(2007)]{gdk+07}
M.~E. {Gonzalez}, R.~{Dib}, V.~M. {Kaspi}, P.~M. {Woods}, C.~R. {Tam}, and
  F.~P. {Gavriil}, \emph{ApJ} (2007), arXiv: 0708.2756 (astro-ph).

  
\end{thebibliography}

\end{document}